\documentclass[12pt]{article}
\usepackage{amsmath}
\usepackage{amssymb}
\usepackage[normalem]{ulem}

\usepackage{graphicx}
\usepackage{epstopdf}
\usepackage{subfig}

\numberwithin{equation}{section}

\usepackage[usenames,dvipsnames]{color}

\begin{document}

\begin{titlepage}
\strut\hfill UMTG--286
\vspace{.5in}
\begin{center}

\LARGE 
Wrapping corrections for non-diagonal boundaries\\
in AdS/CFT\\
\vspace{1in}
\large Zolt\'an Bajnok \footnote{
MTA Lend\"ulet Holographic QFT Group, Wigner Research Centre, H-1525 
Budapest 114, P.O.B. 49, Hungary, bajnok.zoltan@wigner.mta.hu
}
and Rafael I. Nepomechie \footnote{
Physics Department,
P.O. Box 248046, University of Miami, Coral Gables, FL 33124 USA,
nepomechie@physics.miami.edu}\\[0.8in]
\end{center}

\vspace{.5in}

\begin{abstract}
We consider an open string stretched between a $Y=0$
brane and a $Y_\theta=0$ brane. The latter brane is rotated with respect to the 
former by an angle $\theta$, and is described by a non-diagonal 
boundary S-matrix. This system interpolates smoothly
between the $Y-Y$ ($\theta =0$) and the $Y-\bar Y$ ($\theta = \pi /2$)
systems, which are described by diagonal boundary S-matrices.  
We use integrability to compute the energies of 
one-particle states at weak coupling
up to leading wrapping order (4, 6 loops) as a 
function of the angle.
The results for the diagonal cases exactly match with those obtained 
previously.
\end{abstract}

\end{titlepage}

\setcounter{footnote}{0}

\section{Introduction}

The light-cone-gauge worldsheet theory of a free closed type-IIB
superstring on $AdS_{5}\times S^{5}$ is integrable
\cite{Beisert:2010jr}.  Since this string theory is dual
\cite{Maldacena:1997re} to planar ${\cal N}=4$ super Yang-Mills theory
in $3+1$ dimensions, we refer to this worldsheet theory as the
$AdS_{5}/CFT_{4}$ integrable model.  The integrability of this model
with periodic boundary conditions can be exploited to compute the
energies of multiparticle states of the closed string, which coincide
with anomalous dimensions of corresponding single-trace operators in
the dual gauge theory.  For large volumes $L$, the energies are
determined by the asymptotic Bethe ansatz equations
\cite{Beisert:2005fw}, which incorporate all polynomial corrections in
the inverse power of the volume.  The subsequent exponentially-small
finite-size corrections are related \cite{Ambjorn:2005wa} to wrapping
contributions in the gauge theory.  These wrapping corrections are due
to the vacuum polarization effects of bound states on multiparticle
states in the $AdS_{5}/CFT_{4}$ integrable model, which can be
explicitly evaluated \cite{Bajnok:2008bm} and exactly match with
perturbative gauge theory results \cite{Fiamberti:2007rj}.  All
higher-order wrapping corrections can be summed up by the
excited-states thermodynamic Bethe ansatz (TBA) equations
\cite{Gromov:2009bc,Arutyunov:2009ax}, which have a nice reformulation
in terms of the quantum spectral curve \cite{Gromov:2014caa}.

It is also possible to consider the $AdS_{5}/CFT_{4}$ integrable
model on a strip with integrable boundary conditions. This scenario
can be realized by an open string stretched between two maximal giant
gravitons (D-branes). Multiparticle states of the open string correspond
to so-called determinant-like operators in the dual gauge theory.
The $Y-Y$ system, consisting of $Y=0$ branes \cite{Hofman:2007xp}
at both ends of the string, was studied in \cite{Galleas:2009ye,
Bajnok:2010ui, Bajnok:2012xc}.
The integrability of this model was exploited in \cite{Bajnok:2010ui, Bajnok:2012xc}
to compute wrapping corrections of one-particle states. The $Y-\bar{Y}$
system, consisting of a $Y=0$ brane at one end of the string and
a $\bar{Y}=0$ brane at the other end, was subsequently investigated
in \cite{Bajnok:2013wsa}; and wrapping corrections of one-particle
states were again computed. 

We consider here the $Y-Y_{\theta}$ system, consisting of a $Y=0$
brane at one end of the string and a $Y_{\theta}=0$ brane at the other
end, where the latter brane is rotated with respect to the former by
an angle $\theta$.  This system interpolates smoothly between the
$Y-Y$ system ($\theta=0$) and the $Y-\bar{Y}$ system ($\theta=\pi/2$).
We exploit the integrability of this model to compute the leading
wrapping corrections for one-particle states with $L=2$, as
functions of the angle $\theta$.  We verify that these results reduce
for $\theta=0$ and $\theta=\pi/2$ to those obtained previously in
\cite{Bajnok:2010ui, Bajnok:2012xc} and \cite{Bajnok:2013wsa},
respectively.  In principle, it should be possible to confirm these
results from 4-loop and 6-loop computations in the dual gauge theory.

In order to carry out this analysis, it is necessary to know the Bethe-ansatz
expression for the eigenvalues of transfer matrices constructed with
the bulk and boundary worldsheet S-matrices of the $AdS_{5}/CFT_{4}$
integrable model. Since the boundary S-matrix corresponding to the
$Y_{\theta}=0$ brane is generally not diagonal, the problem of determining
the transfer-matrix eigenvalues is nontrivial. Indeed, even for the
much simpler problem of the XXX open spin chain with non-diagonal
boundary terms, a Bethe ansatz solution was obtained only quite recently
using the so-called off-diagonal Bethe ansatz approach \cite{Wang2015}.
With the help of this approach, the Bethe ansatz solution of the AdS/CFT
problem was found in \cite{Zhang:2015fea}. We use that solution here
to formulate the asymptotic Bethe ansatz for the $AdS_{5}/CFT_{4}$
integrable model with non-diagonal boundary 
conditions, and to compute leading wrapping corrections.

The outline of this paper is as follows. In Section 
\ref{sec:ABA}, we first collect all the ingredients needed for the 
computation (S-matrices, transfer matrices, Bethe-ansatz solution, 
etc.), and then calculate the energies of one-particle states with 
$L=2$ at weak coupling up to wrapping order using the asymptotic Bethe ansatz.
In Section \ref{sec:wrapping} we compute the leading wrapping 
corrections for these states, and compare with previous results for 
the diagonal cases. In Section \ref{sec:L1}, we give the 
corresponding results for $L=1$. We conclude in Section 
\ref{sec:discussion} with a brief discussion of our results, and list 
some related open problems.

\section{Asymptotic Bethe ansatz}\label{sec:ABA}

There are two types of finite-size corrections to the energies of
multiparticle states in finite volume.  The leading corrections are
polynomial in the inverse power of the volume and can be accounted for
the momentum quantization of the particles.  These corrections can
be obtained from the Bethe-Yang equation/asymptotic Bethe ansatz,
which implements the periodicity of the wave functions in a very
nontrivial way.  The other corrections are exponentially small in the
volume and have quantum field theoretical origin.  Indeed, these
corrections come from vacuum polarization effects due to the presence
of virtual particles.

In this section, we obtain the finite-size corrections from the asymptotic
Bethe ansatz. We first briefly review the scattering theory of the
$AdS_{5}/CFT_{4}$ integrable model and formulate the boundary Bethe-Yang
equation. We then introduce the relevant transfer matrices, and review
the Bethe-ansatz solution for their eigenvalues. Finally, we use the
asymptotic Bethe ansatz to compute the energies of one-particle states.

\subsection{Fundamental S-matrices}

The $AdS_{5}/CFT_{4}$ integrable model is a (1+1)-dimensional non-relativistic
quantum field theory with a centrally-extended $SU(2|2)\otimes SU(2|2)$
symmetry. The spectrum of this model includes a set of 16 fundamental
particles, which we denote by 
\begin{equation}
|(\alpha,\dot{\alpha})\rangle=|\alpha\rangle\otimes|\dot{\alpha}\rangle\,,\qquad\alpha=1,2,3,4,\qquad\dot{\alpha}=\dot{1},\dot{2},\dot{3},\dot{4,}\label{eq:particlelables}
\end{equation}
where the $SU(2|2)\otimes SU(2|2)$ labels $1,2,\dot{1},\dot{2}$
are bosonic, and $3,4,\dot{3},\dot{4}$ are fermionic. These particles
all have the same energy-momentum dispersion relation 
\begin{equation}
\epsilon(p)=\sqrt{1+16g^{2}\sin^{2}\frac{p}{2}},\qquad g=\frac{\sqrt{\lambda}}{4\pi}\,,\label{eq:epsilon}
\end{equation}
where $\lambda$ is the 't Hooft coupling.

Let us now consider a system of $N$ such particles with momenta $p_{i}$
$(i=1,...,N)$ on a strip of finite length $L$. (Eventually, we shall
restrict to the case $N=1$.) For large $L$, this system can be analyzed
using the bulk and boundary S-matrices of these fundamental particles.
The bulk two-particle S-matrix is given by \cite{Staudacher:2004tk,Beisert:2005tm,Arutyunov:2008zt}
\begin{equation}
\mathbb{S}(p_{1},p_{2})=S_{0}(p_{1},p_{2})\, S(p_{1},p_{2})\otimes\dot{S}(p_{1},p_{2})\,,\label{eq:bulkS}
\end{equation}
whose index structure is given by 
\[
\mathbb{S}_{(\alpha,\dot{\alpha)}(\gamma,\dot{\gamma})}^{(\beta,\dot{\beta)}(\delta,\dot{\delta)}}=S_{0}\, S_{\alpha\gamma}^{\beta\delta}\dot{S}_{\dot{\alpha}\dot{\gamma}}^{\dot{\beta}\dot{\delta}}\,.
\]
Both $S=S_{\alpha\gamma}^{\beta\delta}$ and $\dot{S}=S_{\dot{\alpha}\dot{\gamma}}^{\dot{\beta}\dot{\delta}}$
are given by the graded $16\times16$ matrix in \cite{Arutyunov:2008zt},
which is normalized such that $S_{11}^{11}=\dot{S}_{\dot{1}\dot{1}}^{\dot{1}\dot{1}}=1$,
and the scalar factor $S_{0}$ is given by 
\begin{equation}
S_{0}(p_{1},p_{2})=\frac{x_{1}^{+}+\frac{1}{x_{1}^{+}}-x_{2}^{-}-\frac{1}{x_{2}^{-}}}{x_{1}^{-}+\frac{1}{x_{1}^{-}}-x_{2}^{+}-\frac{1}{x_{2}^{+}}}\frac{x_{1}^{-}}{x_{1}^{+}}\frac{x_{2}^{+}}{x_{2}^{-}}\sigma^{2}(p_{1},p_{2}).\label{eq:S0}
\end{equation}
Here we define $x^{\pm}(p)$ by 
\begin{equation}
x^{\pm}(p)=\frac{1}{4g}(\cot\frac{p}{2}\pm i)(1+\epsilon(p))\,,\label{eq:xpxm}
\end{equation}
and $x_{i}^{\pm}=x^{\pm}(p_{i})$. We shall also make use of the 
rapidity variable
$u$ defined by 
\begin{equation}
x(u)+\frac{1}{x(u)}=\frac{u}{g}\,.\label{eq:udef}
\end{equation}
The dressing factor is given by \cite{Beisert:2006ez,Bajnok:2012bz}
\begin{equation}
\sigma(p_{1},p_{2})=e^{i\Theta(p_{1},p_{2})}\,,\qquad\Theta(p_{1},p_{2})=\chi(x_{1}^{+},x_{2}^{+})+\chi(x_{1}^{-},x_{2}^{-})-\chi(x_{1}^{+},x_{2}^{-})-\chi(x_{1}^{-},x_{2}^{+})\,,
\label{eq:dressing}
\end{equation}
where 
\begin{equation}
\chi(x_{1},x_{2})=-\sum_{r=2}^{\infty}\sum_{s>r}\frac{c_{r,s}(g)}{(r-1)(s-1)}\left[\frac{1}{x_{1}^{r-1}x_{2}^{s-1}}-\frac{1}{x_{1}^{s-1}x_{2}^{r-1}}\right]
\end{equation}
with 
\begin{equation}
c_{r,s}(g)=(r-1)(s-1)2\cos(\tfrac{\pi}{2}(s-r-1))\int_{0}^{\infty}dt\frac{J_{r-1}(2gt)J_{s-1}(2gt)}{t(e^{t}-1)}\,.
\label{eq:crs}
\end{equation}

We assume that the right boundary S-matrix (reflection factor) is
given by 
\begin{equation}
\mathbb{R}^{-}(p)=R_{0}^{-}(p)R^{-}(p)\otimes\dot{R}^{-}(p)\label{eq:BSMminus}
\end{equation}
with \cite{Hofman:2007xp, Chen:2007ec}
\begin{equation}
R_{0}^{-}(p)=-e^{-ip}\sigma(p,-p)\,,\qquad R^{-}(p)=\dot{R}^{-}(p)=\mbox{diag}(e^{-ip/2},-e^{ip/2},1,1)\,.\label{eq:R0Rm}
\end{equation}
This diagonal boundary S-matrix corresponds to a $Y=0$ brane \cite{Hofman:2007xp}.
Let $\mathbb{R}_{\theta}^{-}(p)$ denote the boundary S-matrix
obtained by an angle $\theta$ rotation 
\begin{equation}
\mathbb{R}_{\theta}^{-}(p)=R_{0}^{-}(p)R_{\theta}^{-}(p)\otimes\dot{R}_{\theta}^{-}(p)\,\label{eq:BSMminus2}
\end{equation}
where 
\begin{equation}
R_{\theta}^{-}(p)=O(-\theta)R^{-}(p)\, O(\theta)\,,\qquad O(\theta)=\left(\begin{array}{cccc}
\cos\theta & \sin\theta & 0 & 0\\
-\sin\theta & \cos\theta & 0 & 0\\
0 & 0 & 1 & 0\\
0 & 0 & 0 & 1
\end{array}\right)\,,\label{eq:BSMminus3}
\end{equation}
and same for the dotted indices (i.e., with the same angle $\theta$
for both undotted and dotted factors). In principle we could have a different 
angle $\dot{\theta}$ in $\dot{R}$, but for simplicity we assume 
$\theta=\dot{\theta}$, such that we 
can easily interpolate between the $Y-Y$ ( $\theta=0$) and the 
$Y-\bar Y$ ($\theta=\pi/2$) cases.

We assume that the left boundary
S-matrix is given by \cite{Bajnok:2013wsa} 
\begin{equation}
\mathbb{R}^{+}(p)=\mathbb{R}_{\theta}^{-}(-p)\,.\label{eq:BSMplus}
\end{equation}
This boundary S-matrix, which corresponds to a $Y_{\theta}=0$ brane,
is evidently not diagonal for generic angles.

\subsection{Boundary Bethe-Yang equation}

Our goal is to compute the energies of multiparticle states. For large
$L$, a first approximation to the energy is given by the sum of single-particle
energies 
\begin{equation}
E=\sum_{i=1}^{N}\epsilon(p_{i})\,,\label{eq:totoalenergy}
\end{equation}
where $\epsilon(p)$ is defined in (\ref{eq:epsilon}). Hence, it
is necessary to determine the particle momenta $p_{i}$, which are
quantized for finite $L$. Since the particles have nontrivial scattering,
the quantization condition is given by the boundary Bethe-Yang equation
(see e.g. \cite{Bajnok:2012xc}) 
\begin{equation}
e^{-2ip_{j}L}\prod_{k=j-1}^{1}\mathbb{S}_{jk}(p_{j},p_{k})\,\mathbb{R}_{j}^{-}(p_{j})\prod_{k=1:k\ne j}^{N}\mathbb{S}_{kj}(p_{k},-p_{j})\,\mathbb{R}_{j}^{+}(-p_{j})\,\prod_{k=N}^{j+1}\mathbb{S}_{jk}(p_{j},p_{k})=1\,,\qquad j=1,...,N\,.\label{eq:BBY}
\end{equation}
This condition can be conveniently reformulated in terms of a 
double-row \cite{Sklyanin:1988yz} transfer matrix 
\begin{equation}
\mathbb{D}(p,\{p_{i}\})={\rm tr}_{A}\,\mathbb{S}_{AN}(p,p_{N})...\mathbb{S}_{A1}(p,p_{1})\mathbb{R}_{A}^{-}(p)\mathbb{S}_{1A}(p_{1},-p)...\mathbb{S}_{NA}(p_{N},-p)\tilde{\mathbb{R}}_{A}^{+}(-p)\,,\label{eq:doublerow}
\end{equation}
where the trace is over the auxiliary space denoted here by A, which
is in the fundamental (16-dimensional) representation of $SU(2|2)\otimes SU(2|2)$.
Note that this transfer matrix does not directly depend on the left boundary
S-matrix $\mathbb{R}^{+}(-p)=\mathbb{R}_{\theta}^{-}(p)$, but instead
depends on $\tilde{\mathbb{R}}^{+}(-p)$, which is defined such that
\begin{equation}
\mathbb{R}_{\theta}^{-}(p)_{(\gamma,\dot{\gamma)}}^{(\beta,\dot{\beta)}}=\sum_{\alpha,\dot{\alpha}}\mathbb{S}(p,-p)_{(\alpha,\dot{\alpha)}(\gamma,\dot{\gamma})}^{(\beta,\dot{\beta})(\delta,\dot{\delta})}\tilde{\mathbb{R}}^{+}(-p)_{(\delta,\dot{\delta})}^{(\alpha,\dot{\alpha})}\,.
\end{equation}
The boundary Bethe-Yang equation (\ref{eq:BBY}) now takes the simpler
form 
\begin{equation}
e^{-2ip_{j}L}\mathbb{D}(p_{j},\{p_{i}\})=-1\,,\qquad j=1,...,N\,.
\label{eq:BBY2}
\end{equation}
Using the expressions for the bulk (\ref{eq:bulkS}) and boundary
(\ref{eq:BSMminus}) S-matrices, we can factorize $\mathbb{D}(p,\{p_{i}\})$
into a tensor product of two ``chiral'' $SU(2|2)$ transfer matrices
\begin{equation}
\mathbb{D}(p,\{p_{i}\})=d(p,\{p_{i}\})\, D(p,\{p_{i}\})\otimes\dot{D}(p,\{p_{i}\})\,,
\end{equation}
where 
\begin{equation}
D(p,\{p_{i}\})={\rm tr}_{A}\, 
S_{AN}(p,p_{N})...S_{A1}(p,p_{1})R_{A}^{-}(p)S_{1A}(p_{1},-p)...S_{NA}(p_{N},-p)\tilde{R}_{A}^{+}(-p)\,.\label{eq:Dtransfer}
\end{equation}
The auxiliary space $A$ is now in the fundamental (4-dimensional)
representation of $SU(2|2)$, and similarly for the dotted factor;
moreover, the scalar factor is given by 
\begin{equation}
d(p,\{p_{i}\})=R_{0}^{-}(p)\tilde{R}_{0}^{+}(-p)\prod_{i=1}^{N}S_{0}(p,p_{i})S_{0}(p_{i},-p)\,.\label{eq:dscalarfac}
\end{equation}
We recall \cite{Bajnok:2012xc} that $\tilde{R}^{+}(-p)\propto(-1)^{F}R_{\theta}^{-}(-p)$,
where $F$ is the fermion number, which changes the trace in (\ref{eq:Dtransfer})
to a supertrace. The transfer matrix (\ref{eq:doublerow}) therefore
takes the final form 
\begin{equation}
\mathbb{D}(p,\{p_{i}\})=\tilde{d}(p,\{p_{i}\})\,\tilde{D}(p,\{p_{i}\})\otimes\dot{\tilde{D}}(p,\{p_{i}\})\,,\label{eq:doublerow2}
\end{equation}
where the chiral $SU(2|2)$ transfer matrix $\tilde{D}(p,\{p_{i}\})$
is defined by 
\begin{equation}
\tilde{D}(p,\{p_{i}\})={\rm str}_{A}\, S_{AN}(p,p_{N})...S_{A1}(p,p_{1})R_{A}^{-}(p)S_{1A}(p_{1},-p)...S_{NA}(p_{N},-p)R_{\theta\, A}^{-}(-p)\,,\label{eq:tildeDtransfer}
\end{equation}
and similarly for the dotted factor. The normalization factor is given
by \cite{Bajnok:2012xc} %
\footnote{$\tilde{d}(p,\{p_{i}\})$ must be equal to $d(p,\{p_{i}\})$ up to some scalar
function of $p$. For $N=1$, we see from (\ref{eq:dscalarfac}) that $\tilde{d}(p,p_{1})=g(p)R_{0}^{-}(p)\, S_{0}(p,p_{1})\, S_{0}(p_{1},-p)$
for some function $g(p)$. Evaluating this expression at $p=p_{1}$,
and using (\ref{eq:S0}), (\ref{eq:R0Rm}) and the result $\tilde{d}(p_{1},p_{1})=-e^{-4ip_{1}}\sigma^{2}(p_{1},-p_{1})/\rho_{1}^{2}(p_{1})$
which follows from the boundary Bethe-Yang equation for one $(3,\dot{3})$
particle (see (\ref{eq:BBYN1}) below), we arrive at (\ref{eq:dtildescalarfac}).%
} 
\begin{equation}
\tilde{d}(p,\{p_{i}\})=\frac{e^{-2ip}}{\rho_{1}^{2}(p)}\frac{u^{-}}{u^{+}}\prod_{i=1}^{N}S_{0}(p,p_{i})S_{0}(p_{i},-p)\,.
\label{eq:dtildescalarfac}
\end{equation}
where 
\begin{equation}
\rho_{1}(p)=\frac{(1+(x^{-})^{2})(x^{-}+x^{+})}{2x^{+}(1+x^{-}x^{+})}\,.
\label{eq:rho1}
\end{equation}

For later reference, we recall here that there exists an infinite
hierarchy of commuting transfer matrices 
$\mathbb{D}_{a,s}(p,\{p_{i}\})$
\begin{equation}
\left[\mathbb{D}_{a,s}(p,\{p_{i}\})\,,\mathbb{D}_{a',s'}(p',\{p_{i}\})\right]=0\,,
\end{equation}
defined as in (\ref{eq:doublerow}) except with the auxiliary space
in a rectangular representation $(a,s)$ of $SU(2|2)\otimes SU(2|2)$,
such that $\mathbb{D}(p,\{p_{i}\})\equiv\mathbb{D}_{1,1}(p,\{p_{i}\})$.
These transfer matrices satisfy the Hirota equation 
\begin{equation}
\mathbb{D}_{a,s}^{+}\,\mathbb{D}_{a,s}^{-}=\mathbb{D}_{a+1,s}\,\mathbb{D}_{a-1,s}+\mathbb{D}_{a,s+1}\,\mathbb{D}_{a,s-1}\,,
\end{equation}
where $f^{\pm}(u)=f(u\pm\frac{i}{2})$. As in (\ref{eq:doublerow2}),
we can express $\mathbb{D}_{a,s}$ in terms of corresponding chiral
$SU(2|2)$ transfer matrices 
\begin{equation}
\mathbb{D}_{a,s}(p,\{p_{i}\})=\tilde{d}_{a,s}(p,\{p_{i}\})\,\tilde{D}_{a,s}(p,\{p_{i}\})\otimes\dot{\tilde{D}}_{a,s}(p,\{p_{i}\})\,.
\label{eq:doublerow2gen}
\end{equation}

\subsection{Bethe ansatz}

In order to determine the momenta $p_{i}$ using the boundary Bethe-Yang
equation (\ref{eq:BBY2}), it is necessary to first determine the
eigenvalues of $\mathbb{D}(p,\{p_{i}\})$. In view of (\ref{eq:doublerow2}),
the problem in turn reduces to determining the eigenvalues of the
chiral transfer matrix $\tilde{D}(p,\{p_{i}\})$ (\ref{eq:tildeDtransfer}).
The latter problem is nontrivial due to the fact that the boundary
S-matrix $R_{\theta}^{-}(p)$ (\ref{eq:BSMminus3}) is not diagonal.
Nevertheless, with the help of the so-called off-diagonal Bethe ansatz
approach \cite{Wang2015}, this problem was recently solved in \cite{Zhang:2015fea}.
The result for the eigenvalues of $\tilde{D}(p,\{p_{i}\})$ (which,
by abuse of notation, we denote in the same way) is given by \cite{Zhang:2015fea}%
\footnote{We compensate for the fact that the definition of $g$ in \cite{Zhang:2015fea}
differs by a factor 2 from the one used here. Indeed, there $g=\sqrt{\lambda}/(2\pi)$,
c.f. (\ref{eq:epsilon}). Moreover, we change notation $B_{1}R_{3}\mapsto{\cal R}_{1}\,,R_{1}B_{3}\mapsto{\cal B}_{1}$.%
} 
\begin{eqnarray}
 &  & \tilde{D}(p,\{p_{i}\})=e^{i(N-M+1)p}\frac{{\cal R}^{(+)-}}{{\cal R}^{(+)+}}\rho_1 
 \Bigg\{-\frac{{\cal R}^{(-)-}}{{\cal R}^{(+)-}}\frac{{\cal R}_{1}^{+}}{{\cal R}_{1}^{-}}-\frac{u^+}{u^-}
 \frac{{\cal B}^{(+)+}}{{\cal B}^{(-)+}}\frac{{\cal B}_{1}^{-}}{{\cal B}_{1}^{+}}\label{tildeDeig}\\
 &  & \quad\quad+\frac{1}{2}\biggl(1+\frac{u^+}{u^-}\biggr)\Bigg[\frac{u^{-}}{u}\frac{{\cal R}_{1}^{+}}{{\cal R}_{1}^{-}}\frac{Q_{2}^{--}}{Q_{2}}+\frac{u^{+}}{u}\frac{{\cal B}_{1}^{-}}{{\cal B}_{1}^{+}}\frac{Q_{2}^{++}}{Q_{2}}-4\sin^{2}\theta\frac{Q_{1}^{-}{\cal R}_{1}^{+}}{Q_{2}{\cal R}_{1}^{-}}\Bigg]\Bigg\}\,,\nonumber 
\end{eqnarray}
%\begin{eqnarray}
%\lefteqn{\tilde{D}(p,\{p_{i}\})=e^{i(N-M+1)p}\frac{1}{{\cal R}^{(+)+}{\cal B}^{(-)+}}\Bigg\{-\rho_{1}{\cal R}^{(-)-}{\cal B}^{(-)+}\frac{{\cal R}_{1}^{+}}{{\cal R}_{1}^{-}}-\rho_{2}{\cal R}^{(+)-}{\cal B}^{(+)+}\frac{{\cal B}_{1}^{-}}{{\cal B}_{1}^{+}}}\label{tildeDeig}\\
% &  & \quad\quad+\tfrac{1}{2}(\rho_{1}+\rho_{2}){\cal R}^{(+)-}{\cal B}^{(-)+}\Bigg[\frac{u^{-}}{u}\frac{{\cal R}_{1}^{+}}{{\cal R}_{1}^{-}}\frac{Q_{2}^{--}}{Q_{2}}+\frac{u^{+}}{u}\frac{{\cal B}_{1}^{-}}{{\cal B}_{1}^{+}}\frac{Q_{2}^{++}}{Q_{2}}-4\sin^{2}\theta\frac{Q_{1}^{-}{\cal R}_{1}^{+}}{Q_{2}{\cal R}_{1}^{-}}\Bigg]\Bigg\}\,,\nonumber 
%\end{eqnarray}
where
\begin{equation}
{\cal R}^{(\pm)}(p)=\prod_{i=1}^{N}\left(x(p)-x^{\mp}(p_{i})\right)\left(x(p)+x^{\pm}(p_{i})\right)\,,\quad
{\cal R}_{1}(p)=\prod_{j=1}^{M}\left(x(p)-y_{j}\right)\left(x(p)+y_{j}\right)\,,\label{eq:RBdefs}
\end{equation}
and their $\mathcal{B}$ analogues are obtained by changing $x(p)$
to $1/x(p)$: 
\begin{equation}
{\cal 
B}^{(\pm)}(p)=\prod_{i=1}^{N}\left(\frac{1}{x(p)}-x^{\mp}(p_{i})\right)\left(\frac{1}{x(p)}+x^{\pm}(p_{i})\right)\,,\quad 
{\cal 
B}_{1}(p)=\prod_{j=1}^{M}\left(\frac{1}{x(p)}-y_{j}\right)\left(\frac{1}{x(p)}+y_{j}\right)\,.
\label{eq:RBdefs}
\end{equation}
We shall often use the abbreviation $x^{\pm}_{i} = x^{\pm}(p_{i})$.
The $Q$-functions are 
\begin{equation}
Q_{1}(u)=\prod_{j=1}^{M}(u-v_{j})(u+v_{j})\,,\quad  
Q_{2}(u)=\prod_{j=1}^{M}(u-w_{j})(u+w_{j})\,, 
\label{Q1Q2}
\end{equation}
where $v_{j}=g(y_{j}+\tfrac{1}{y_{j}})$. Finally, $\rho_{1}$ is given by 
(\ref{eq:rho1}).
%, and $\rho_{2}$ is given by 
%\begin{equation}
%\rho_{2}(p)=\frac{x^{-}(1+(x^{+})^{2})(x^{-}+x^{+})}{2(x^{+})^{2}(1+x^{-}x^{+})}\,.
%\label{eq:rho2}
%\end{equation}

The corresponding Bethe equations for the Bethe roots $\{y_{1},\ldots,y_{M}\}$
and $\{w_{1},\ldots,w_{M}\}$ are 
\begin{eqnarray}
 &  & \frac{{\cal R}^{(-)}}{{\cal R}^{(+)}}\frac{Q_{2}^+}{Q_{2}^{-}}\Bigg\vert_{x(p)=y_{j}}=1\,,\qquad j=1,\ldots,M\,,\label{BAE1a}\\
 &  & \left[\tfrac{u^{-}}{u}Q_{1}^{+}Q_{2}^{--}+\tfrac{u^{+}}{u}Q_{1}^{-}Q_{2}^{++}-4\sin^{2}\theta\, Q_{1}^{+}Q_{1}^{-}\right]\Bigg\vert_{u=w_{k}}=0\,,\quad k=1,\cdots,M\,,\label{BAE2a}
\end{eqnarray}
For a given value of $N$, the possible values of $M$ are $0,1,\ldots N$.
Notice the presence of the ``inhomogeneous'' term in (\ref{tildeDeig})
that is proportional to $\sin^{2}\theta$, which is absent for the
diagonal ($\theta=0$) case \cite{Bajnok:2012xc}.

In order to compute the L\"uscher corrections, we shall also need the
corresponding result for all the antisymmetric representations $\tilde{D}_{a,1}(p,\{p_{i}\})$.
A generating functional for these transfer-matrix eigenvalues was
proposed in \cite{Zhang:2015fea}, which we now briefly recall. We
begin by rewriting the eigenvalue result (\ref{tildeDeig}) for $\tilde{D}=\tilde{D}_{1,1}$
as 
\begin{equation}
\tilde{D}_{1,1}=h\,\hat{D}_{1,1}\,,\qquad\hat{D}_{1,1}=-A-B+G+H+C\,,
\label{eq:tildeD11}
\end{equation}
where $h$ is a normalization factor 
\begin{equation}
h=\rho_{1}\left(\frac{x^{+}}{x^{-}}\right)^{N-M+1}\frac{{\cal 
R}^{(+)-}}{{\cal R}^{(+)+}}\,.
\label{eq:h}
\end{equation}
Furthermore,
\begin{equation}
A=\frac{{\cal R}^{(-)-}}{{\cal R}^{(+)-}}\frac{{\cal R}_{1}^{+}}{{\cal R}_{1}^{-}}\,,\quad B=\frac{u^{+}}{u^{-}}\frac{B^{(+)+}}{B^{(-)+}}\frac{{\cal B}_{1}^{-}}{{\cal B}_{1}^{+}}\,,\quad G=\frac{{\cal R}_{1}^{+}}{{\cal R}_{1}^{-}}\frac{Q_{2}^{--}}{Q_{2}}\,,\quad H=\frac{u^{+}}{u^{-}}\frac{{\cal B}_{1}^{-}}{{\cal B}_{1}^{+}}\frac{Q_{2}^{++}}{Q_{2}}\,,\label{eq:ABCGH}
\end{equation}
and the $\theta$-dependent term is 
\begin{equation}
C=-2\sin^{2}\theta\left(1+\frac{u^{+}}{u^{-}}\right)\frac{Q_{1}^{-}{\cal R}_{1}^{+}}{Q_{2}{\cal R}_{1}^{-}}\,.
\end{equation}
The proposed generating functional for antisymmetric representations
is given by \cite{Zhang:2015fea} 
\begin{eqnarray}
W^{-1} & = & (1-{\cal D}A{\cal D})^{-1}\left[1-{\cal D}(G+H+C){\cal D}+{\cal D}G{\cal D}^{2}H{\cal D}\right](1-{\cal D}B{\cal D})^{-1}\nonumber \\
 & = & \sum_{a=0}^{\infty}(-1)^{a}{\cal D}^{a}\,\hat{D}_{a,1}\,{\cal D}^{a}\,,\label{genfunc}
\end{eqnarray}
where ${\cal D}=e^{-\frac{i}{2}\partial_{u}}$ implying ${\cal D}f=f^{-}{\cal D}$,
with 
\begin{equation}
\tilde{D}_{a,1}=h^{[a-1]}h^{[a-3]}\cdots h^{[3-a]}h^{[1-a]}\,\hat{D}_{a,1}\,,
\end{equation}
where $f^{[\pm n]}=f(u\pm\frac{in}{2})$.

\subsection{One-particle states}

For simplicity, we henceforth focus on the case $N=1$. (The case $N=0$,
corresponding to the vacuum state, was considered in \cite{Bajnok:2013wsa}.)
For this case, the boundary Bethe-Yang equation (\ref{eq:BBY}) reduces
to 
\begin{eqnarray}
1 & = & 
e^{-2ip_{1}L}\Lambda(p_{1})=e^{-2ip_{1}L}R_{0}(p_{1})^{2}\lambda_{i}(p_{1})\dot{\lambda}_{j}(p_{1})\,,
\label{eq:BBYN1}
\end{eqnarray}
where $\Lambda(p_{1})$ denotes an eigenvalue of 
$\mathbb{R}^{-}(p_{1}) \,\mathbb{R}^{+}(-p_{1}) =\mathbb{R}^{-}(p_{1}) \,\mathbb{R}_{\theta}^{-}(p_{1})\,$.
Recalling (\ref{eq:BSMminus}) and (\ref{eq:BSMminus2}), we see that
there are 16 such eigenvalues, which are given (up to the factor 
$R_{0}(p_{1})^{2}$) by
the products of the 4 eigenvalues of $ R^{-}(p_{1}) \,R_{\theta}^{-}(p_{1})$,
denoted by $\lambda_{i}(p_{1})$, and the 4 eigenvalues of $\dot{R}^{-}(p_{1})\, \dot{R}_{\theta}^{-}(p_{1})$,
denoted by $\dot{\lambda}_{j}(p_{1})$.
We restrict our attention throughout this paper to the 4 symmetric 
$\dot{\lambda}_{j}=\lambda_{i}$
cases. The two eigenvalues corresponding to the bosonic subspace are
\begin{equation}
\lambda_{1,2}=\cos p_{1}\,\cos^{2}\theta-\sin^{2}\theta\pm i\sqrt{1-(\sin^{2}\theta-\cos p_{1}\,\cos^{2}\theta)^{2}}\,,
\label{eq:lambda12}
\end{equation}
while those in the fermionic subspace are $\lambda_{3,4}=1$. 

Since the Bethe-Yang equation (\ref{eq:BBYN1}) can also be written in terms 
of the transfer matrix as in (\ref{eq:BBY2}), we must have $\Lambda(p_{1}) = 
-\mathbb{D}(p_{1},p_{1})$. Indeed, the eigenvalues (\ref{eq:lambda12}) can be recovered from the
Bethe-ansatz result for the transfer-matrix eigenvalue
(\ref{tildeDeig}) as follows: the fermionic eigenvalues are described
by $N=1$ and $M=0$ as 
\begin{equation}
\tilde{D}(p_{1},p_{1})\big\vert_{M=0}=e^{ip_{1}}\rho_{1}(p_{1}) 
\end{equation}
(note that ${\cal R}^{(+)-}(p_{1})=0$), while the bosonic 
eigenvalues are described by $N=M=1$ as
\begin{equation}
\tilde{D}(p_{1},p_{1})\big\vert_{M=1}=\rho_{1}(p_{1}) 
\frac{\mathcal{R}_{1}^{+}(p_{1})}{\mathcal{R}_{1}^{-}(p_{1})} \,.
\end{equation}
Their ratio (which must coincide with $\lambda_{1,2}/\lambda_{3,4} 
= \lambda_{1,2}$) is therefore given by
\begin{equation}
\frac{\tilde{D}(p_{1},p_{1})\big\vert_{M=1}}{\tilde{D}(p_{1},p_{1})\big\vert_{M=0}} = 
e^{-ip_{1}}\frac{\mathcal{R}_{1}^{+}(p_{1})}{\mathcal{R}_{1}^{-}(p_{1})} 
= e^{-ip_{1}}\frac{(x_{1}^{+}-y_{1})(x_{1}^{+}+y_{1})}{(x_{1}^{-}-y_{1})(x_{1}^{-}+y_{1})} \,,
\label{eq:deig1par}
\end{equation}
where the ``magnonic'' Bethe roots $y_{1}$ 
(or $v_{1}=g(y_{1} + \frac{1}{y_{1}})$) and $w_{1}$ are still to be determined from the 
Bethe equations (\ref{BAE1a}), (\ref{BAE2a}) in terms of $p_{1}$.
The $Q$-functions (\ref{Q1Q2}) simplify for $N=M=1$ to 
\begin{equation}
Q_{1}(u)=(u-v_{1})(u+v_{1})\,, \qquad Q_{2}(u)=(u-w_{1})(u+w_{1}) \,.
\end{equation}
The Bethe equation (\ref{BAE2a}) expresses $w_{1}^{2}$ in terms of $v_{1}$ as 
\begin{equation}
w_{1}^{2}=v_{1}^{2}-v_{1}\cot\theta-\frac{1}{4}\,, \qquad 
w_{1}^{2}=v_{1}^{2}+v_{1}\cot\theta-\frac{1}{4}\,.
\label{eq:vw}
\end{equation}
As the second equation can be obtained from the first by changing $\theta\to-\theta$
or $v_{1}\to-v_{1}$, we focus on the first equation. It implies that
\begin{equation}
\frac{Q_{2}^{+}(v_{1})}{Q_{2}^{-}(v_{1})}=e^{2i\theta} \,,
\end{equation}
which simplifies the other Bethe equation (\ref{BAE1a}) to 
\begin{equation}
\frac{(y_{1}-x_{1}^{+})(y_{1}+x_{1}^{-})}{(y_{1}-x_{1}^{-})(y_{1}+x_{1}^{+})}e^{2i\theta}=1 \,.
\end{equation}
This quadratic equation has two solutions for $y_{1}$ 
\begin{equation}
y_{1}=x_{1}^{-}\frac{e^{ip_{1}/2}}{\sin\theta}\left(\cos\theta\sin\frac{p_{1}}{2}\mp\sqrt{1-\cos^{2}\frac{p_{1}}{2}\cos^{2}\theta}\right) \,.
\end{equation}
Plugging these two solutions back into (\ref{eq:deig1par}) we recover
the two eigenvalues (\ref{eq:lambda12}). Let us note that taking
the second solution in eq. (\ref{eq:vw}), i.e. changing 
$\theta\to-\theta$, alters the sign of $y_{1}$ but does not change 
the expression (\ref{eq:deig1par}).

\subsection{Energies from the asymptotic Bethe ansatz at weak coupling}\label{sec:ABAenergies}

The dressing phase appears in the boundary Bethe-Yang equations
(\ref{eq:BBYN1}) (recall that $R_{0}^{-}(p)$ is given by 
(\ref{eq:R0Rm})), which prevents us from solving these equations explicitly.
However, in order to compare with gauge-theory calculations, 
only the weak-coupling (small $g$) expansion is needed. We 
now develop this expansion for $L=2$, since
several results that can be used as checks are already available for 
this case.
Restricting to symmetric states ($\dot{\lambda}_{j}=\lambda_{i}$) and taking the square root
of the boundary Bethe-Yang equations (\ref{eq:BBYN1}), we obtain
\begin{equation}
1=\gamma e^{-3ip_{1}}\sigma(p_{1},-p_{1})\, \lambda_{i}(p_{1}) \,.
\label{eq:sqrtBBYN1}
\end{equation}
We keep track of the square root sign ambiguity by introducing 
$\gamma=\pm 1$.
For the computations that follow, it turns out to be advantageous to work
with the rapidity variable $u$ (\ref{eq:udef}) instead of the momentum. They are related
as 
\begin{equation}
u(p)=\frac{1}{2}\cot\frac{p}{2}\,\epsilon(p) \,.
\end{equation}
Our tactic is to first expand $u_{1}$ at weak coupling as 
\begin{equation}
u_{1}=u_{1,0} + g^{2}\, u_{1,1} + g^{4}\, u_{1,2}    +\dots \,,
\end{equation}
and to substitute these results into the Bethe-Yang equation 
(\ref{eq:sqrtBBYN1}), expanding also the dressing factor $\sigma(p_{1},-p_{1})$ using 
(\ref{eq:dressing})-(\ref{eq:crs}) and 
\begin{equation}
x(u)=\frac{u}{2g}+\sqrt{\frac{u}{2g}+1}\sqrt{\frac{u}{2g}-1} \,.
\label{xu}
\end{equation}
We then solve the resulting equation
order by order in $g$. Once $u_{1}$ is known up to the required order,
we substitute the result into the energy formula
\begin{equation}
\epsilon(u)=1+2ig\left(\frac{1}{x^{+}}-\frac{1}{x^{-}}\right)\,, 
\label{eq:epsilon2}
\end{equation}
which we also expand. Since we are interested in the leading wrapping
correction, we expand the energy up to that order. We now summarize 
our results for all the possible choices of $\lambda_{i}$ and $\gamma$ 
in (\ref{eq:sqrtBBYN1}):

\begin{enumerate}
    
\item $\lambda_{3}=\lambda_{4}=1$ and $\gamma=+1$: 
the leading weak-coupling result for $u_{1}$ is given by
\begin{equation}
u_{1,0}=\frac{1}{2\sqrt{3}} \,,
\label{eq:u1case1}
\end{equation}
which corresponds to $p_{1,0}=2\pi/3$, and 
which gets modified up to $g^{6}$ as 
\begin{equation}
u_{1}=u_{1,0}\left(1+6g^{2}-18g^{4}+108g^{6}+24g^{6}\zeta_{3}+\dots\right) \,.
\end{equation}
This leads to the energy
\begin{equation}
E_{1}=1+6g^{2}-18g^{4}+108g^{6}-18(45+4\zeta_{3})g^{8} \,.
\label{eq:E1}
\end{equation}

\item $\lambda_{3}=\lambda_{4}=1$ and $\gamma=-1$:
we find
\begin{equation}
u_{1,0}=\frac{\sqrt{3}}{2} \,,
\label{eq:u1case2}
\end{equation}
which corresponds to $p_{1,0}=\pi/3$, and 
which gets modified up to $g^{6}$ as 
\begin{equation}
u_{1}=u_{1,0}\left(1+2g^{2}-2g^{4}+4g^{6}+\tfrac{8}{3}g^{6}\zeta_{3}+\dots\right) \,.
\end{equation}
This leads to the energy
\begin{equation}
E_{2}=1+2g^{2}-2g^{4}+4g^{6}-2(5+4\zeta_{3})g^{8} \,.
\label{eq:E2}
\end{equation}

\item $\lambda_{1,2}$ and $\gamma=+1$: the leading order gives the
relation
\begin{equation}
\cos(2\theta)=\frac{17-56u_{1,0}^{2}+16u_{1,0}^{4}}{(1+4u_{1,0}^{2})^{2}} \,.
\end{equation}
Solving this relation for $u_{1,0}^{2}$, we obtain 
\begin{equation}
u_{1,0}^{2}=-\frac{1}{4}+\frac{1}{1\pm\cos\theta} \,,
\label{u1gammaplus}
\end{equation}
where $\lambda_{1}$ and $\lambda_{2}$ are compatible with the upper 
and lower signs, respectively.
The corrections to the rapidity up to $g^{10}$ can be expressed as  
\begin{align}
u_{1} = & u_{1,0} 
\left(1+8\tilde{g}^{2}-32\tilde{g}^{4}+256\tilde{g}^{6}-2560\tilde{g}^{8}+28672\tilde{g}^{10}\right)+ \\
& u_{1,0}^{-1} (1 - 12u_{1,0}^2 ) \left (
-4 {\tilde g}^6 (1 + 4 u_{1,0}^2) \zeta_3 +   {\tilde g}^8 \left ( 16(1 + 20 u_{1,0}^2) \zeta_3 + 
  40 (1 + 4 u_{1,0}^2)^2 \zeta_5 \right ) \right . \nonumber \\
 & \left . - {\tilde g}^{10} \left ( 64 (1 + 84 u_{1,0}^2) \zeta_3 + 
 224 (1 + 16 u_{1,0}^2 + 48 u_{1,0}^4) \zeta_5 + 
  420 (1 + 4 u_{1,0}^2)^3 \zeta_7 \right ) \right) \nonumber \, ,
\end{align}
where $\tilde{g}^{2}=g^{2}/(1+4u_{1,0}^{2})$. 
The corresponding energy is 
\begin{align}
E_{3}= & 1+8\tilde{g}^{2}-32\tilde{g}^{4}+256\tilde{g}^{6}-2560\tilde{g}^{8}+28672\tilde{g}^{10}-344064\tilde{g}^{12}\nonumber \\
 & 
 +(1-12u_{1,0}^{2})\left[256\tilde{g}^{8}\zeta_{3}-2560\tilde{g}^{10}(2\zeta_{3}+(1+4u_{1,0}^{2})\zeta_{5})+\right.\nonumber \\
 & 
 \left.\qquad\qquad\qquad+768\tilde{g}^{12}(112\zeta_{3}+8(9+28u_{1,0}^{2})\zeta_{5}+35(1+4u_{1,0}^{2})^{2}\zeta_{7}\right] \,.
\label{eq:E3}
\end{align}
\item $\lambda_{1,2}$ and $\gamma=-1$: the leading-order equation
gives 
\begin{equation}
\cos(2\theta)=\frac{1}{2u_{1,0}^{2}}-\frac{17-56u_{1,0}^{2}+16u_{1,0}^{4}}{(1+4u_{1,0}^{2})^{2}} \,.
\label{u1gammaminus}
\end{equation}
Since the correction to $u_{1,0}$ is quite complicated, we refrain 
from displaying the result. The corresponding energy correction
takes the form
\begin{align}
E_4= &  1+8\tilde{g}^{2}-32\tilde{g}^{4}+256\tilde{g}^{6}-2560\tilde{g}^{8}+28672\tilde{g}^{10}-344064\tilde{g}^{12}\nonumber \\
&- \frac {32768 u_{1,0}^4 (3 - 4u_{1,0}^2)}{1 + 24 u_{1,0}^2 - 48 u_{1,0}^4} 
\left ( {\tilde g}^8 \zeta_3  -10 {\tilde g}^{10} (2 \zeta_3 +(1 + 4 u_{1,0}^2 )\zeta_5 ) \right . \nonumber \\
& \quad \qquad \left .+ {\tilde g}^{12} \left ( 336 \zeta_3 + 24 (9 + 28 u_{1,0}^2) \zeta_5 + 
  105 (1 + 4 u_{1,0}^2)^2 \zeta_7 \right) \right ) \,.
\label{eq:E4}
\end{align}
\end{enumerate}

For case 3, we see from  (\ref{u1gammaplus}) that there are two 
solutions for $u_{1,0}^{2}$ in terms of $\cos\theta$. Since 
\begin{equation}
u_{1,0} = \frac{1}{2}\cot(\tfrac{p_{1,0}}{2})\,,
\end{equation}
where $p_{1,0}$ is the weak-coupling limit of the momentum $p_{1}$,
it follows that $p_{1,0}$ and $\theta$ are related in a simple 
manner
\begin{equation}
\cos p_{1,0} = \sin^{2}\tfrac{\theta}{2}\,, \qquad \cos p_{1,0} = 
\cos^{2}\tfrac{\theta}{2}\,.
\end{equation}
These two solutions are plotted in Fig 1(a).

For case 4, it follows from (\ref{u1gammaminus}) that there are three
solutions for $u_{1,0}^{2}$. The corresponding momenta $p_{1,0}$ as 
functions of angle are plotted in Fig 1(b).

\begin{figure}
\centering
\subfloat[]{\includegraphics[width=7.5cm]{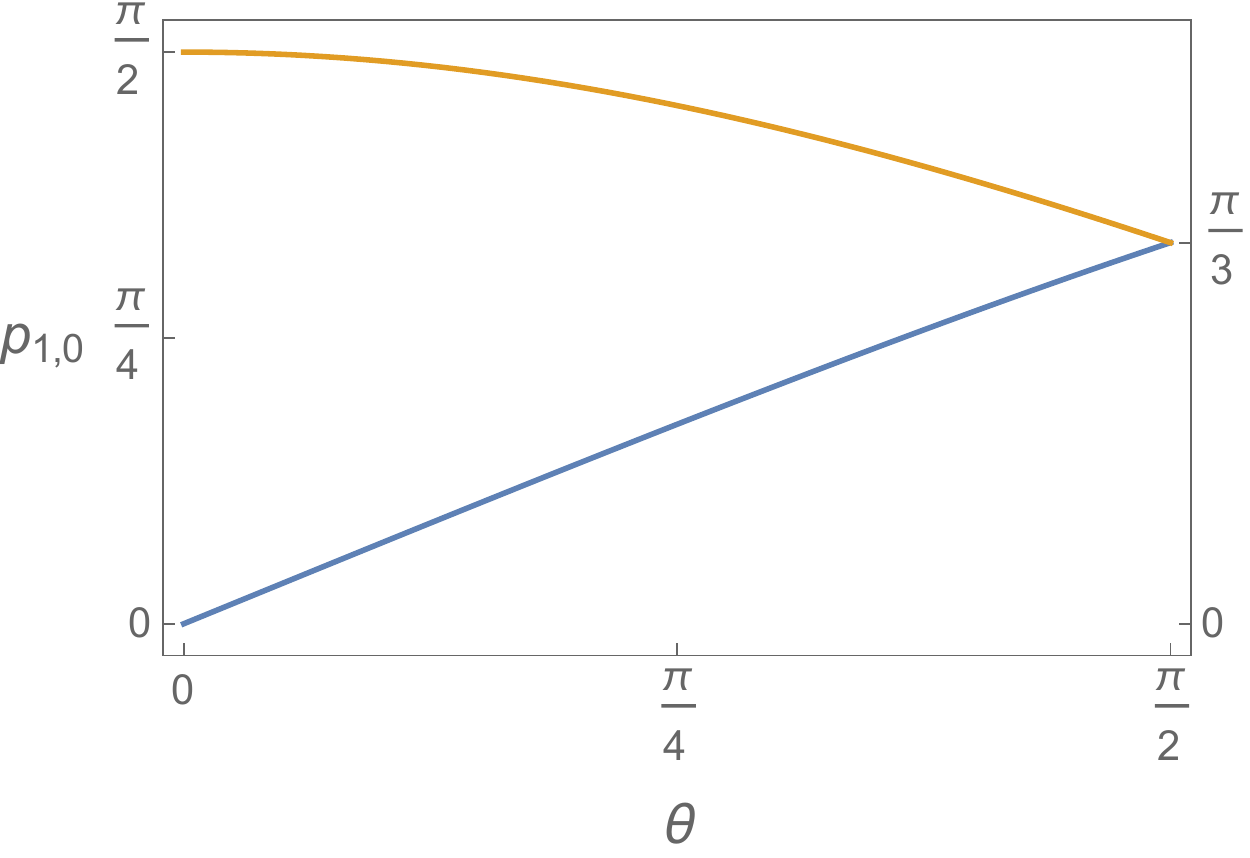}}
\hspace{1cm}
\subfloat[]{\includegraphics[width=7.5cm]{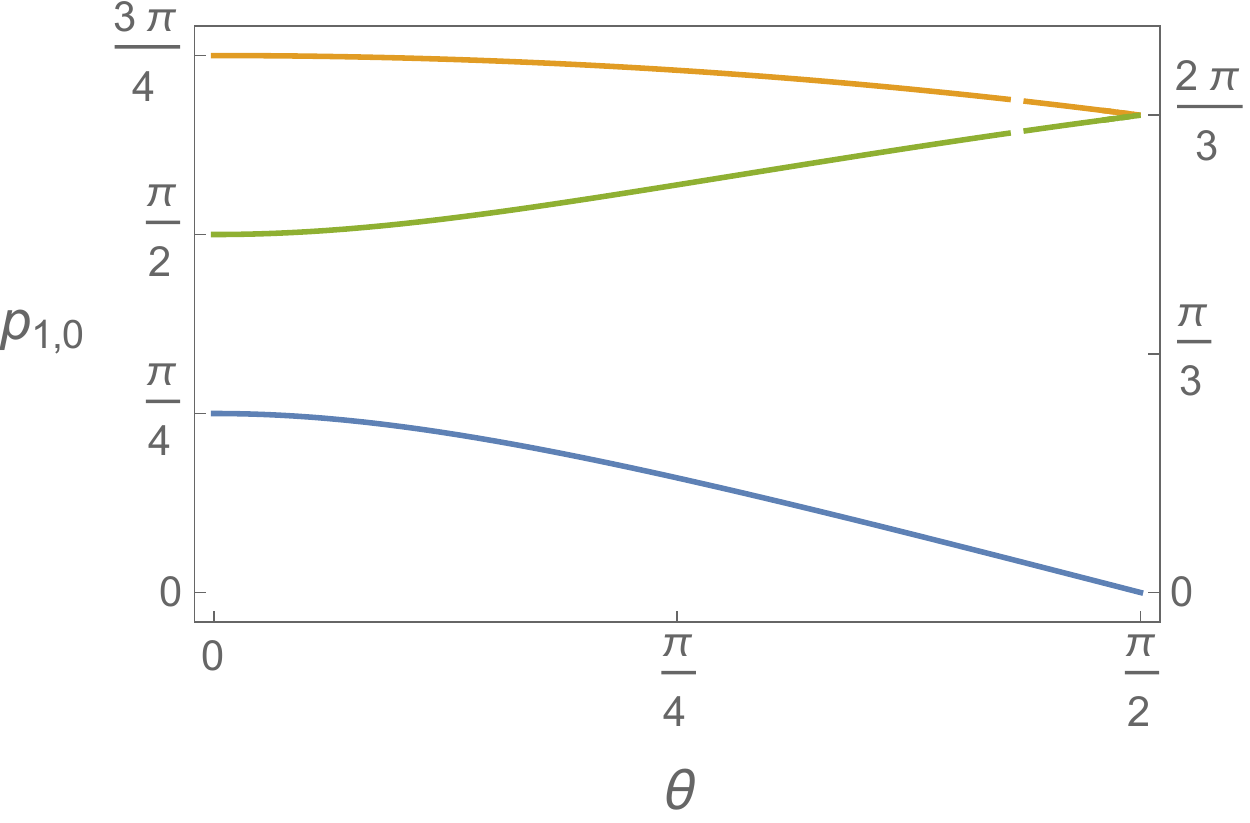}}
\caption{$p_{1,0}$ versus $\theta$ for (a) case 3 (b) case 4} 
\label{fig:plots}
\end{figure}

\subsection{Weak-coupling expansion of the magnonic Bethe roots}\label{sec:weakcoupling}

We will calculate in Section \ref{sec:wrapping} the leading wrapping corrections to
the energies $E_{i}$ computed above. For cases 3 and 4, we will need the leading weak-coupling
expressions for the Bethe roots $v_{1}$ and $w_{1}$ in terms of $u_{1,0}$.
We set 
\begin{equation}
v_{1}  = v_{1,0} + O(g^{2}) \,, \qquad  
w_{1}  = w_{1,0} + O(g^{2}) \,,
\end{equation}
and we note that $y_{1} \sim \tfrac{1}{g}v_{1,0}$.
Let us now see how the Bethe equations simplify for small $g$. We
begin by introducing the $Q$-function corresponding to $u_1$
\begin{equation}
Q(u)=(u-u_1)(u+u_1) \,.
\end{equation}
In the weak-coupling limit, $x^{\pm} \sim \frac{1}{g}(u \pm 
\frac{i}{2})$, and therefore $g^2 \mathcal{R}^{(\pm)}=Q^{\pm}$. The
first Bethe equation (\ref{BAE1a}) therefore simplifies to 
\begin{equation}
\left.\frac{Q^{-}Q_{2}^{+}}{Q^{+}Q_{2}^{-}}\right|_{u=v_{1,0}}=1 \,,
\end{equation}
which implies that at leading order the $w_{1}$ root is
the same as $u_{1}$: 
\begin{equation}
w_{1,0}=u_{1,0}\,, \qquad Q_{2}(u)=Q(u)+O(g^2) \,.
\label{eq:w1equ1}
\end{equation}
The result (\ref{eq:deig1par}) from the
transfer-matrix eigenvalue also simplifies
\begin{equation}
\lambda_{1,2} = e^{-ip_{1}} 
\left.\frac{Q_{1}^{+}}{Q_{1}^{-}}\right|_{u=u_{1,0}}   \,,   
\end{equation}    
and leads to the following expression for the (square root of the) 
Bethe-Yang equation (\ref{eq:sqrtBBYN1}) with $\lambda_{1,2}$ at weak 
coupling\footnote{For $L\ne 2$, the right-hand-side of (\ref{eq:BetheYang}) changes to
$\gamma\left(\frac{u_{1,0}+\frac{i}{2}}{u_{1,0}-\frac{i}{2}}\right)^{L+2}$.}
\begin{equation}
\left.\frac{Q_{1}^{+}}{Q_{1}^{-}}\right|_{u=u_{1,0}}=\gamma\left(\frac{u_{1,0}+\frac{i}{2}}{u_{1,0}-\frac{i}{2}}\right)^{4} \,.
\label{eq:BetheYang}
\end{equation}
This equation can be used to express $v_{1,0}$ in terms of $u_{1,0}$,
which -- when plugged back into (\ref{eq:vw}) -- determines $u_{1,0}$ in
terms of $\theta$, or the other way around. We find two
solutions for $v_{1,0}^{2}$ in terms of $u_{1,0}^{2}$: 
\begin{eqnarray}
v_{1,0}^{2} &=&\frac{(1+4u_{1,0}^{2})^{2}}{32u_{1,0}^{2}-8} \qquad\qquad\qquad\quad
(\gamma=+1) \,, 
\label{eq:gammaplus} \\
\qquad 
v_{1,0}^{2}&=&-\frac{(1+4u_{1,0}^{2})^{2}(1-4u_{1,0}^{2})}{4-96u_{1,0}^{2}+64u_{1,0}^{4}} \qquad 
(\gamma=-1) \,.
\label{eq:gammaminus}
\end{eqnarray}
The former solution corresponds to $\gamma=+1$ and is related to case 3 analyzed
above; while the latter solution corresponds to $\gamma=-1$ and case 4.

The first relation (\ref{eq:gammaplus}), when combined 
with (\ref{eq:vw}), recovers the result (\ref{u1gammaplus}) for $u_{1,0}^{2}$ in
terms of $\cos\theta$; and correspondingly, $v_{1,0}=\pm\csc\theta$.
The second relation (\ref{eq:gammaminus}), when combined 
with (\ref{eq:vw}), recovers the result (\ref{u1gammaminus}).
We prefer to use the variable $u_{1,0}$ instead of $\theta$ 
as both the Bethe-Yang energy and the wrapping correction can be 
expressed in terms of $u_{1,0}$ in a unified way. That is, we have a 
single expression for the two cases at $\gamma=1$, and another expression for 
the three cases at $\gamma=-1$.

We are finally ready to calculate the wrapping corrections to the 
one-particle states.

\section{Leading wrapping corrections}\label{sec:wrapping}

The leading finite-size correction of multiparticle states on the
strip has been proposed in \cite{Bajnok:2010ui}. It expresses the
energy corrections in terms of the double-row transfer-matrix 
eigenvalue as:
\begin{equation}
\Delta 
E=-\sum_{a=1}^{\infty}\int_{0}^{\infty}\frac{dq}{2\pi}\mathbb{D}_{a,1}(q,p_{1})e^{-2\tilde{\epsilon}_{a}(q)L} \,.
\label{eq:wrapping}
\end{equation}
From (\ref{eq:doublerow2gen}) and (\ref{eq:tildeD11}) we have
\begin{equation}
\mathbb{D}_{a,1}(q,p_{1})=f_{a,1}(q,p_{1})\,\hat{D}_{a,1}(q,p_{1})^{2}\,,
\label{eq:wrapdoublerow}
\end{equation}
which must be evaluated for the mirror momenta $q$. The scalar part
can be obtained from fusion 
\begin{equation}
f_{a,1}=f^{[a-1]}f^{[a-3]}\dots f^{[3-a]}f^{[1-a]}\,,
\label{eq:fusedf}
\end{equation}
where $f$ is given by
\begin{equation}
f= \tilde{d}(q,p_{1})\, h^{2} =
S_{0}(q,p_{1})S_{0}(p_{1},-q)\frac{u^{-}}{u^{+}}\left(\frac{{\cal 
R}^{(+)-}}{{\cal 
R}^{(+)+}}\right)^{2}\left(\frac{x^{+}}{x^{-}}\right)^{2(N-M)} \,,
\end{equation}
as follows from (\ref{eq:dtildescalarfac}) and (\ref{eq:h}).
In calculating the fusion of $a$ particles to get the mirror 
antisymmetric boundstate, we must take
the first $a-1$ particles in the ``string'' kinematics, i.e.  use 
\eqref{xu} with $u=q/2$; and take only the last $a^{th}$
particle in the mirror kinematics \cite{Bajnok:2008bm,Arutyunov:2007tc}, where we have
\begin{equation}
e^{-\tilde{\epsilon}_{a}(q)}=\frac{x^{[-a]}(q)}{x^{[+a]}(q)}\,,
\qquad x^{[\pm 
a]}(q)=\frac{q+ia}{4g}\left(\sqrt{1+\frac{16g^{2}}{q^{2}+a^{2}}}\pm1\right) \,.
\label{xq}
\end{equation}
We calculate the boundstate transfer-matrix eigenvalues from the generating
functional (\ref{genfunc}) as 
\begin{equation}
(-1)^{a}\hat{D}_{a,1}=\sum_{j=0}^{a}A^{(j)}B^{(a-j)}-\sum_{j=0}^{a-1}A^{(j)}J^{[a-1-2j]}B^{(a-j-1)}
+\sum_{j=0}^{a-2}A^{(j)}G^{[a-1-2j]}H^{[a-3-2j]}B^{(a-j-2)} 
\label{eq:fromgenfunc}
\end{equation}
where $J=G+H+C$, and
\begin{equation}
A^{(j)}=A^{[a-1]}A^{[a-3]}\dots A^{[a+1-2j]}=\frac{{\cal R}^{(-)[a-2]}}{{\cal R}^{(+)[a-2]}}\frac{{\cal R}^{(-)[a-4]}}{{\cal R}^{(+)[a-4]}}\dots
\frac{{\cal R}^{(-)[a-2j]}}{{\cal R}^{(+)[a-2j]}}\frac{{\cal 
R}_{1}^{[a]}}{{\cal R}_{1}^{[a-2j]}}\,,
\label{eq:A(j)}
\end{equation}
together with 
\begin{equation}
B^{(k)}=B^{[2k-1-a]}\dots B^{[3-a]} B^{[1-a]}=\frac{u^{[2k-a]}}{u^{[-a]}}\frac{{\cal B}^{(+)[2k-a]}}{{\cal B}^{(-)[2k-a]}}\dots
\frac{{\cal B}^{(+)[4-a]}}{{\cal B}^{(-)[4-a]}}\frac{{\cal 
B}^{(+)[2-a]}}{{\cal B}^{(-)[2-a]}}\frac{{\cal B}_{1}^{[-a]}}{{\cal 
B}_{1}^{[2k-a]}} \,.
\label{eq:B(j)}
\end{equation}
In the following we specialize the above expressions for the four cases 
that we analyzed in Section \ref{sec:weakcoupling}, and calculate 
their weak-coupling limits.

\subsection{Wrapping corrections to $\lambda_{3}$}\label{sec:wrappingeasy}

For the $(3,\dot{3})$ particle there are no magnons ($M=0$), thus 
\begin{equation}
Q_{1}=Q_{2}=\mathcal{R}_{1}=\mathcal{B}_{1}=1 \,.
\end{equation}
In the weak-coupling limit $g^{2} \mathcal{R}^{(\pm)}=Q^{\pm}$, 
hence
\begin{equation}
A^{(k)}=\frac{Q^{[a-2k-1]}}{Q^{[a-1]}}\quad\mbox{for}\,\quad  k<a\,,
\qquad 
A^{(a)}=\frac{Q^{[1-a]}}{Q^{[a-1]}}\,, 
\qquad B^{(k)}=\frac{u^{[2k-a]}}{u^{[-a]}}\,,
\end{equation}
and 
\begin{equation}
H^{[k]}=\frac{u^{[k+1]}}{u^{[k-1]}}\,,\qquad 
G^{[k]}=1 \,, \qquad
G^{[k]}H^{[k-2]}=\frac{u^{[k-1]}}{u^{[k-3]}}\,, \qquad 
C^{[k]}=-\sin^{2}\theta\frac{4u^{[k]}}{u^{[k-1]}}
\,.
\end{equation}
Performing the sums in (\ref{eq:fromgenfunc}), we obtain the 
following result for the transfer-matrix part 
\begin{equation}
\hat{D}_{a,1}(q,u_{1,0})=(-1)^{a+1}\,\frac{aq\sin^{2}\theta(a^{2}-1-q^{2}+4u_{1,0}^{2})}{(q-ia)Q^{[a-1]}} \,,
\end{equation}
where we have used the leading-order rapidity $u_{1,0}$ instead of the momentum
$p_{1}$. The weak-coupling limit of the scalar part (\ref{eq:fusedf}) gives
\begin{equation}
f_{a,1}(q,u_{1,0})=\frac{Q^{[a-1]}(u_{1,0}^{2}+\frac{1}{4})^{2}}{Q^{[-a-1]}Q^{[1-a]}Q^{[a+1]}}\frac{q-ia}{q+ia} \,.
\end{equation}
The weak-coupling limit of $\mathbb{D}_{a,1}$ 
(\ref{eq:wrapdoublerow}) is therefore given by
\begin{equation}
\mathbb{D}_{a,1}(q,u_{1,0}) =\frac{a^{2}q^{2}\sin^{4}\theta\left(\, a^{2}-1-q^{2}+4u_{1,0}^{2}\right)^{2}}{Q^{[-a-1]}Q^{[1-a]}Q^{[a-1]}Q^{[1+a]}}
\frac{(u_{1,0}^{2}+\frac{1}{4})^{2}}{(q^{2}+a^{2})} \,.
\end{equation}
The exponential part is simply 
\begin{equation}
e^{-2\tilde{\epsilon}_{a}L}=\left(\frac{4g^{2}}{q^{2}+a^{2}}\right)^{4} \,,
\label{eq:expart}
\end{equation}
where we have taken $L=2$.
As the integrand in (\ref{eq:wrapping}) is symmetric in $q$ we extend
the integral to the whole line and evaluate it by residues. On the
upper half-plane there is a kinematical pole at $q=ia$ and four dynamical
poles at $q=i(a\pm1+2u_{1,0})$ and at $q=i(a\pm1-2u_{1,0})$. We find
that the contributions from the dynamical poles at $q=i(a+1+2u_{1,0})$
(and similarly for the dynamical poles at $q=i(a+1-2u_{1,0})$)
coming from two consecutive values of $a$ cancel provided that $u_{1,0}$ satisfies
the Bethe-Yang equations, i.e. it is either  
$1/(2\sqrt{3})$ (\ref{eq:u1case1}) or
$\sqrt{3}/2$  (\ref{eq:u1case2}). The contributions coming from the kinematical
pole can be summed up, and we obtain the following results: 
\begin{enumerate}
\item For $u_{1,0}=\frac{1}{2\sqrt{3}}$ we obtain
\begin{equation}
\Delta E_{1}=2g^{8}\sin^{4}\theta(3\zeta_{3}-5\zeta_{5}) \,.
\end{equation}

\item For $u_{1,0}=\frac{\sqrt{3}}{2}$ we obtain 
\begin{equation}
\Delta E_{2}=-2g^{8}\sin^{4}\theta(\zeta_{3}+5\zeta_{5}) \,.
\end{equation}
\end{enumerate}
The corresponding energies from the asymptotic Bethe 
ansatz are given by (\ref{eq:E1}) and  (\ref{eq:E2}), respectively.

\subsection{Wrapping correction to $\lambda_{1,2}$}

We now consider the more complicated cases. 
One must be careful in calculating the weak-coupling expansion of the 
eigenvalues $\hat{D}_{a,1}$  in the mirror kinematics. As already explained, the first $a-1$ particles are 
in the ``string'' kinematics, while only the last $a^{th}$
particle is in the mirror kinematics. This implies the following weak-coupling behavior: 
\begin{equation}
x^{[j]}=\frac{q+ij}{2g}+\dots \quad\mbox{for}\,\quad j=a,\dots,1-a\,,\qquad x^{[-a]}=\frac{2g}{q-ia}+\dots
\end{equation}
and will introduce a difference in the expansion of the $\mathcal{R}$
and $\mathcal{B}$ functions depending on their shifts: 
\begin{equation}
g^2\mathcal{R}^{(\pm)[k]}=Q^{[k\pm1]}+\dots \quad\mbox{for}\,\quad 
k>-a\,,\qquad g^2\mathcal{R}^{(\pm)[-a]}=-(u_{1,0}^{2}+\frac{1}{4}) \,,
\end{equation}
and
\begin{equation}
g^2\mathcal{B}^{(\pm)[k]}=-(u_{1,0}^{2}+\frac{1}{4})+\dots 
\quad\mbox{for}\,\quad 
k>-a\,,\qquad g^2 \mathcal{B}^{(\pm)[-a]}=Q^{[-a\pm1]} \,.
\end{equation}
Similarly for $\mathcal{R}_{1}$ and $\mathcal{B}_{1}$ we obtain 
\begin{equation}
g^2{\cal R}_{1}^{[k]}=Q_{1}^{[k]}+ \dots \quad\mbox{for}\,\quad 
k>-a\,,\qquad g^2 {\cal R}_{1}^{[-a]}=-v_{1,0}^{2} \,,
\end{equation}
and
\begin{equation}
g^2 {\cal B}_{1}^{[k]}=-v_{1,0}^{2}+\dots \quad\mbox{for}\,\quad 
k>-a\,,\qquad g^2 {\cal B}_{1}^{[-a]}=Q_{1}^{[-a]} \,.
\end{equation}
Substituting these results into the expression for $A^{(k)}$ 
(\ref{eq:A(j)}), we arrive at 
\begin{equation}
A^{(k)}=\frac{Q^{[a-2k-1]}}{Q^{[a-1]}}\frac{Q_{1}^{[a]}}{Q_{1}^{[a-2k]}} \quad\mbox{for}\,\quad  k<a\,,
\qquad 
A^{(a)}=\frac{Q^{[1-a]}}{Q^{[a-1]}}\frac{Q_{1}^{[a]}}{(-v_{1,0}^{2})} 
\,.
\end{equation}
For $B^{(k)}$ (\ref{eq:B(j)}) we do not have this problem, as it is always evaluated
in the last mirror kinematics
\begin{equation}
B^{(k)}=\frac{u^{[2k-a]}}{u^{[-a]}}\frac{Q_{1}^{[-a]}}{(-v_{1,0}^{2})} 
\,.
\end{equation}
For the other terms in the transfer-matrix eigenvalues we obtain
\begin{equation}
G^{[k]}=\frac{Q_{1}^{[k+1]}}{Q_{1}^{[k-1]}}\frac{Q_{2}^{[k-2]}}{Q_{2}^{[k]}}\quad\mbox{for}\,\quad k>1-a\,, 
\qquad 
G^{[1-a]}=\frac{Q_{1}^{[2-a]}Q_{2}^{[-a-1]}}{(-v_{1,0}^{2})Q_{2}^{[1-a]}}\,,
\end{equation}
\begin{equation}
H^{[k]}=\frac{u^{[k+1]}}{u^{[k-1]}}\frac{Q_{2}^{[k+2]}}{Q_{2}^{[k]}}\quad\mbox{for}\,\quad k>1-a\,,
\qquad 
H^{[1-a]}=\frac{u^{[2-a]}}{u^{[-a]}}\frac{Q_{1}^{[-a]}}{(-v_{1,0}^{2})}\frac{Q_{2}^{[3-a]}}{Q_{2}^{[1-a]}}\,,
\end{equation}
\begin{equation}
C^{[k]}=-2\sin^{2}\theta\,\frac{2u^{[k]}}{u^{[k-1]}}\frac{Q_{1}^{[k+1]}}{Q_{2}^{[k]}}\quad\mbox{for}\,\quad k>1-a\,,
\qquad 
C^{[1-a]}=-2\sin^{2}\theta\,\frac{2u^{[1-a]}}{u^{[-a]}}\frac{Q_{1}^{[2-a]}}{Q_{2}^{[1-a]}}\frac{Q_{1}^{[-a]}}{(-v_{1,0}^{2})} \,.
\end{equation}
Finally, 
\begin{equation}
G^{[k]}H^{[k-2]}=\frac{Q_{1}^{[k+1]}}{Q_{1}^{[k-1]}}\frac{u^{[k-1]}}{u^{[k-3]}}\quad\mbox{for}\,\quad k>3-a\,,
\qquad 
G^{[3-a]}H^{[1-a]}=\frac{Q_{1}^{[4-a]}}{Q_{1}^{[2-a]}}\frac{Q_{1}^{[-a]}}{(-v_{1,0})^{2}}\frac{u^{[2-a]}}{u^{[-a]}} \,.
\end{equation}
We recall (\ref{eq:w1equ1}) that $Q_{2}=Q$ at leading order, which simplifies the sums 
in (\ref{eq:fromgenfunc}), leading to a remarkably compact expression
\begin{equation}
\hat{D}_{a,1}(q,u_{1,0})=(-1)^{a+1}\frac{aq(\sin^{2}\theta\,(q^{4}+q^{2}(2a^{2}-8v_{1,0}^{2})+(a^{2}+4v_{1,0}^{2})^{2})-16v_{1,0}^{2})}{4v_{1,0}^{2}(q-ia)Q^{[a-1]}} \,.
\end{equation}
Here $v_{1,0}$ is not independent of $u_{1,0}$ as they can be related
either by (\ref{eq:vw}) or by (\ref{eq:BetheYang}). Using 
(\ref{eq:vw}),
the explicit $\theta$ dependence can be factored out as
\begin{equation}
\hat{D}_{a,1}(q,u_{1,0})=(-1)^{a+1}\frac{aq\sin^{2}\theta\,((a^{2}+q^{2})^{2}+8v_{1,0}^{2}(a^{2}-q^{2}+4u_{1,0}^{2}-1)-(1+4u_{1,0}^{2})^{2})}{4v_{1,0}^{2}(q-ia)Q^{[a-1]}} \,.
\end{equation}
The weak-coupling limit of the scalar part (\ref{eq:fusedf}) is 
\begin{equation}
f_{a,1}(q,u_{1,0})=\frac{Q^{[a-1]}(u_{1,0}^{2}+\frac{1}{4})^{2}}{Q^{[-a-1]}Q^{[1-a]}Q^{[a+1]}}\left(\frac{4g^{2}}{q^{2}+a^{2}}\right)^{2}\frac{q-ia}{q+ia} \,.
\end{equation}
The full contribution of $\mathbb{D}_{a,1}$ (\ref{eq:wrapdoublerow}) 
is therefore
\begin{eqnarray}
\mathbb{D}_{a,1} 
&=&\frac{a^{2}q^{2}\sin^{4}\theta\,\left((a^{2}+q^{2})^{2}+8v_{1,0}^{2}(a^{2}-q^{2}+4u_{1,0}^{2}-1)-(1+4u_{1,0}^{2})^{2}\right)^{2}}{Q^{[-a-1]}Q^{[1-a]}Q^{[a-1]}Q^{[1+a]}} \nonumber\\
&&\times\left(\frac{4g^{2}}{q^{2}+a^{2}}\right)^{2}\frac{(u_{1,0}^{2}+\frac{1}{4})^{2}}{16v_{1,0}^{4}(q^{2}+a^{2})} \,.
\label{finalDD}
\end{eqnarray}
The exponential part is again given by (\ref{eq:expart}).

Since the integrand (\ref{eq:wrapping}) is again symmetric in $q$, we extend the integral
to the whole line and evaluate it by residues. On the upper 
half-plane we find the same poles that we found in Section \ref{sec:wrappingeasy}
for the $\lambda_{3}$ case.
We also find that the contributions from the dynamical poles at $q=i(a+1+2u_{1,0})$
(and similarly for the dynamical poles at $q=i(a+1-2u_{1,0})$)
coming from two consecutive values of $a$ cancel provided $u_{1,0}$ and $v_{1,0}$
are related by the Bethe-Yang equation i.e. (\ref{eq:gammaplus}) or (\ref{eq:gammaminus}). 
Summing up only the contributions from the kinematical residues we 
find the following results: 
\begin{enumerate}
\setcounter{enumi}{2}
\item For $\lambda_{1,2}$ with $\gamma=+1$ and (\ref{eq:gammaplus}),
\begin{equation}
\Delta 
E_{3}=-49152 {\tilde g}^{12}(1 - 4 u_{1,0}^2)^2\zeta_{5} 
\label{deltaE3}
\,.
\end{equation}

\item For $\lambda_{1,2}$ with $\gamma=-1$ and (\ref{eq:gammaminus}),
\begin{equation}
\Delta E_{4}= 3 {\tilde g}^{12} (1 - 24 u_{1,0}^2 + 16 u_{1,0}^4)^2 u_{1,0}^{-4}(256 u_{1,0}^{2} \zeta_5 - 
   7 (1 + 4 u_{1,0}^2)^4 \zeta_9) \,.
\label{deltaE4}
\end{equation}
% They are understood that the relations between $u_{1,0}$ and $\theta$
% coming from the leading order Bethe-Yang equation have to be satisfied. 
\end{enumerate}
The corresponding energies from the asymptotic Bethe 
ansatz are given by (\ref{eq:E3}) and  (\ref{eq:E4}), respectively.

Let us now compare these results with those obtained previously for 
the two diagonal cases:
\begin{description}

\item[$\theta=0$:] For the $Y-Y$ case with $\gamma=+1$,  
(\ref{u1gammaplus}) implies that there are two solutions 
$u_{1,0}=\infty\,, 1/2$,  
corresponding to $p_{1,0} = 0, \pi/2$, as can be seen from Fig 1(a).
From (\ref{deltaE3}) it follows that,  up to $g^{12}$, there are no wrapping 
corrections, $\Delta E_{3} = 0$. For the $Y-Y$ case with 
$\gamma=-1$, we find from  (\ref{u1gammaminus}) 
that there are three solutions $u_{1,0}= (\sqrt{2}\pm 1)/2\,, 1/2$,
corresponding to $p_{1,0} = \pi/4\,, 3\pi/4\,, \pi/2$, as can be seen from Fig 1(b).
From (\ref{deltaE4}) it follows that,  up to $g^{12}$, there are no wrapping 
corrections for $p_{1,0} = \pi/4\,, 3\pi/4$, and
\begin{equation}
\Delta E_{4}\Big\vert_{p_{1,0} = 
\pi/2} =768 g^{12}\zeta_{5} -1344 g^{12}\zeta_{9} \,,
\end{equation}
which agrees with (4.26) in \cite{Bajnok:2012xc}.

\item[$\theta=\pi/2$:] For the $Y-\bar{Y}$ case with $\gamma=+1$, 
(\ref{u1gammaplus}) implies that there is just one solution 
$u_{1,0}=\sqrt{3}/2$,  
corresponding to $p_{1,0} = \pi/3$, as can be seen from Fig 1(a).
From (\ref{deltaE3}) we obtain the wrapping correction
\begin{equation}
\Delta E_{3}\Big\vert_{p_{1,0} = 
\pi/3} = -48 g^{12}\zeta_{5} \,,
\end{equation}
which agrees with (D.15) in \cite{Bajnok:2013wsa}. For the $Y-\bar{Y}$ case with 
$\gamma=-1$, we find from  (\ref{u1gammaminus})
that there are two solutions $u_{1,0}= \infty\,, 1/(2\sqrt{3})$,
corresponding to $p_{1,0} = 0\,, 2\pi/3$, as can be seen from Fig 1(b).
From (\ref{deltaE4}) we obtain the wrapping correction
\begin{equation}
\Delta E_{4}\Big\vert_{p_{1,0} = 
2\pi/3} = 1296 g^{12}\zeta_{5} -1344 g^{12}\zeta_{9}\,,
\end{equation}
which agrees with (D.16) in \cite{Bajnok:2013wsa}.

\end{description}
 In short, the results (\ref{deltaE3}) and
(\ref{deltaE4}) for the wrapping corrections are in complete
agreement with those obtained previously for $\theta=0$ and
$\theta=\pi/2$.  While the boundary S-matrix $R_{\theta}^{-}(p)$
(\ref{eq:BSMminus3}) is diagonal for both of these angles, the
``extra'' term in the Bethe-ansatz solution (\ref{tildeDeig}) does
{\em not} vanish for $\theta=\pi/2$.  \footnote{For a discussion of
this point, see \cite{Zhang:2015fea}.} Hence, the agreement at
$\theta=\pi/2$ provides strong support for the Bethe-ansatz solution
(\ref{tildeDeig}) and for the corresponding generating functional
(\ref{genfunc}).

\section{Results for $L=1$}\label{sec:L1}

In this section we analyze the energies of the states related to the
$\lambda_{1,2}$ eigenvalues for $L=1$ up to the leading wrapping
correction. Although the wrapping correction for the vacuum state
at $L=1$ seems to be divergent \cite{Bajnok:2013wsa}, our calculation formally makes sense
also for this case. 

The boundary Bethe-Yang equation for $L=1$ symmetric states ($\lambda_i=\dot{\lambda}_i$) takes the form 
(c.f. (\ref{eq:sqrtBBYN1}))
\begin{equation}
1=\gamma e^{-2ip_{1}}\sigma(p_{1},-p_{1})\lambda_{i}(p_{1})\,, 
\qquad\gamma=\pm 1\,. \label{eq:sqrtBBYL1}
\end{equation}
Depending on the sign of $\gamma$, we find the following results.
\begin{description}
\item[$\gamma=1$]: The Bethe-Yang equation (\ref{eq:sqrtBBYL1}) at leading order gives the following
relation between the angle $\theta$ and the rapidity $u_{1,0}$:
\begin{equation}
\cos(2\theta)=1+\frac{1}{2u_{1,0}^{2}}-\frac{8}{1+4u_{1,0}^{2}} \,.
\end{equation}
The higher-order corrections are 
\begin{equation}
u_{1}=u_{1,0}\left(1+8\tilde{g}^{2}-32\tilde{g}^{4}+256\tilde{g}^{6}\right)
+1024\tilde{g}^{6}\frac{u_{1,0}^{3}(1+4u_{1,0}^{2})}{1+12u_{1,0}^{2}}\zeta_{3} \,,
\end{equation}
such that the energy up to order $g^{8}$ is 
\begin{equation}
E=1+8\tilde{g}^{2}-32\tilde{g}^{4}+256\tilde{g}^{6}-2560\tilde{g}^{8}
-65536\tilde{g}^{8}\frac{u_{1,0}^{4}}{1+12u_{1,0}^{2}}\zeta_{3} \,.
\end{equation}
We recall that  $\tilde{g}^{2}=g^{2}/(1+4u_{1,0}^{2})$. 
At this order, wrapping starts to contribute as 
\begin{equation}
\Delta E=32\tilde{g}^{8}u_{1,0}^{-2}(1-12u_{1,0}^{2})^{2}\zeta_{3}
-\frac{5}{2}\tilde{g}^{8}u_{1,0}^{-4}(1-8u_{1,0}^{2}-48u_{1,0}^{4})^{2}\zeta_{5} \,.
\end{equation}
For this case, we have 
$v_{1}^{2}=(1+4u_{1,0}^{2})^{2}/(48u_{1,0}^{2}-4)$.

\item[$\gamma=-1$]: The leading-order rapidity $u_{1,0}$ can be expressed
in terms of $\theta$ as 
\begin{equation}
\cos(2\theta)=-1+\frac{8}{1+4u_{1,0}^{2}} \,,
\end{equation}
which can be easily inverted to give 
$u_{1,0}^{2}=\tan^{2}\theta+\frac{3}{4}$.
Its corrections are given by 
\begin{equation}
u_{1}=u_{1,0}\left(1+8\tilde{g}^{2}-32\tilde{g}^{4}+256\tilde{g}^{6}\right)
-16\tilde{g}^{6}u_{1,0}^{-1}(1-16u_{1,0}^{4})\zeta_{3} \,,
\end{equation}
which lead to the energy
\begin{equation}
E=1+8\tilde{g}^{2}-32\tilde{g}^{4}+256\tilde{g}^{6}-2560\tilde{g}^{8}
+1024\tilde{g}^{8}(1-4u_{1,0}^{2})\zeta_{3} \,.
\end{equation}
The leading wrapping correction to this state is 
\begin{equation}
\Delta E=-128\tilde{g}^{8}(3-4u_{1,0}^{2})^{2}\zeta_{3}-40\tilde{g}^{8}
(3+8u_{1,0}^{2}-16u_{1,0}^{4})^{2}\zeta_{5} \,.
\end{equation}
\end{description}
For this case, we have $v_{1}^{2}=\frac{5}{4}+u_{1,0}^{2}+4/(4u_{1,0}^{2}-3)$.

\section{Discussion}\label{sec:discussion}

We have analyzed the leading wrapping corrections for one-particle
states in the $AdS_{5}/CFT_{4}$ integrable model on the strip with
non-diagonal boundary conditions at one end.  This boundary system
describes the excitations of an open string stretched between a $Y=0$
brane and a rotated $Y_\theta=0$ brane, which interpolates smoothly
between the $Y-Y$ ($\theta =0$) and the $Y-\bar Y$ ($\theta = \pi /2$)
systems.  Our analysis has two novel features: the use of a Bethe
ansatz solution with an ``inhomogeneous'' boundary-dependent term
(\ref{tildeDeig}), (\ref{BAE2a}); and the presence of a pair of
momentum-dependent magnonic rapidities\footnote{
In the work \cite{Sfondrini:2011rr} the authors also considered the 
wrapping correction of magnonic states, but only with $y$-type roots.} 
($v_{1}$ and $w_{1}$, which
depend on the momentum through a continuous parameter $\theta$) to
determine the boundstate transfer-matrix eigenvalues 
$\mathbb{D}_{a,1}$, which are needed to obtain the leading exponential finite-size
corrections. Due to the unusual generating 
functional (\ref{genfunc}) and the presence of the magnonic Bethe 
roots, the intermediate expressions are quite complicated. 
Nevertheless, the final expression (\ref{finalDD}) for $\mathbb{D}_{a,1}$
is remarkably compact.
Our results provide the evolution of the energies of all
excitations for sizes $L=1$ and $L=2$ up to the leading wrapping order from
$\theta =0$ to $\theta= \pi/2$, and reproduce the available limiting
cases.  Interestingly, the energies exhibit a smooth behavior, even
though the ground state develops a tachyonic instability
\cite{Bajnok:2013wsa}.

The $AdS_{5}/CFT_{4}$ integrable model admits other interesting 
non-diagonal boundary conditions, for which vacuum wrapping corrections 
were calculated in \cite{Correa:2009mz, Correa:2012hh,Drukker:2012de,Bajnok:2013sya}.
It would be interesting to extend those analyses for one-particle states
and compare the structure of the results with our findings.

In calculating the leading wrapping correction, it was enough to take
into account the effect of vacuum polarization on the energy.  This is
due to the fact that the dispersion relation (\ref{eq:epsilon}), 
(\ref{eq:epsilon2})
contains the coupling constant.  At the next-to leading order wrapping
correction, the effect of vacuum polarization on the boundary Bethe-Yang
equation should also be taken into account \cite{Bajnok:2010ui}.  It
would be very interesting to derive these corrections for non-diagonal
boundaries.

In order to sum up all (leading as well as sub-leading) wrapping
corrections, one should derive the corresponding TBA equations.  These
equations could also shed some light on the tachyonic instability, as
by changing the angle smoothly one could switch from the stable $Y-Y$
system to the unstable $Y-\bar Y$ system.  The TBA equations would be
the first step towards deriving a more compact formulation of exact
finite-size energies, and could lead to a non-diagonal boundary
generalization of the quantum spectral curve \cite{Gromov:2014caa}.

\section*{Acknowledgments}

ZB thanks the University of Miami for its hospitality, and 
acknowledges the support of a Lend\"ulet Grant. 
The work of RN was supported in part by the National Science
Foundation under Grant PHY-1212337, and by a Cooper fellowship.

% \bibliographystyle{utphys}
% \bibliography{refs}

\providecommand{\href}[2]{#2}\begingroup\raggedright\endgroup

\end{document}